
\input harvmac
\def \D {{\cal D}} \def \bt {{\bar \theta }}\def \bp {
\Phi }
 \def \t {\theta}\def \ra {\rightarrow}
\def \eq#1 {\eqno{(#1)}}

\def \a {\alpha}
\def \ga {\alpha}
\def \b {\beta}

\def \gb {\beta}
\def \ga {\alpha}
\def \o {\omega}

\def \gp {\phi}
\def \p {\phi}

\def\gg {\gamma}

\def \r {\rho}
\def \d {\delta}
\def \l {\lambda}
\def \m {\mu}
\def \g {\gamma}
\def \n {\nu}
\def \gd {\delta}
\def \gs {\sigma}

\def \bgb {\bar \beta }
\def \gij {g_{ij}}
\def \Gmn {G_{\mu \nu}}
\def \fourth {{1\over 4}}
\def \third {{1\over 3}}
\def \e#1 {{\rm e}^{#1}}
\def \const {{\rm const }}

\def \vp {\varphi}
\def \ggij {{g_{ij}}}
\def \dg  {{\dot g}}
\def \ddg  {{ \ddot g}}
\def \df   {{\dot f}}
\def \ddf  {{\ddot f}}
\def \hg {{\hat g}}
\def \B {{\hat B}}

\def \hgg {{\hat \g}}
\def \hR {{\hat R}}
\def \tg {{\tilde g}}
\def \ha { { 1\over 2 }}
\def \dpp   {{\dot \phi }}
\def \ddp  {{\ddot \phi}}

\def \eq#1 {\eqno{(#1)}}

\def \a {\alpha}
\def \ga {\alpha}
\def \del {{\partial}}
\def \b {\beta}

\def \gb {\beta}
\def \ga {\alpha}
\def \o {\omega}

\def \gp {\phi}
\def \p {\phi}

\def\gg {\gamma}

\def \r {\rho}
\def \d {\delta}
\def \l {\lambda}
\def \m {\mu}
\def \g {\gamma}
\def \n {\nu}
\def \gd {\delta}
\def \gs {\sigma}
\def \t {\theta}
\def \bgb {\bar \beta }
\def \gij {g_{ij}}
\def \Gmn {G_{\mu \nu}}
\def \fourth {{1\over 4}}
\def \third {{1\over 3}}
\def \e#1 {{\rm e}^{#1}}
\def \const {{\rm const }}

\def \vp {\varphi}
\def \ggij {{g_{ij}}}
\def \dg  {{\dot g}}
\def \ddg  {{ \ddot g}}
\def \df   {{\dot f}}
\def \ddf  {{\ddot f}}
\def \hg {{\hat g}}
\def \B {{\hat B}}

\def \hgg {{\hat \g}}
\def \hR {{\hat R}}
\def \tg {{\tilde g}}
\def \ha { { 1\over 2 }}
\def \dpp   {{\dot \phi }}
\def \ddp  {{\ddot \phi}}

\def\np {  Nucl. Phys. }
\def \pl { Phys. Lett. }
\def \mpl { Mod. Phys. Lett. }
\def \prl { Phys. Rev. Lett. }
\def \pr  { Phys. Rev. }
\def \ijmp  { Int.J.Mod.Phys. }
\def \ap { Ann. Phys.  }
\def \cmp  { Commun.Math.Phys. }

\def\e#1{{\rm e}^{^{\textstyle#1}}}

\def\darr#1{\raise1.5ex\hbox{$\leftrightarrow$}\mkern-16.5mu #1}
\def\ha{{1\over2}}
\def\half{{\textstyle{1\over2}}} 
\def\roughly#1{\raise.3ex\hbox{$#1$\kern-.75em\lower1ex\hbox{$\sim$}}}

\baselineskip=16pt plus 2pt minus 2pt

\Title{\vbox{\baselineskip16pt\hbox{Imperial/TP/92-93/7 }\hbox{hep-th/9211061}
}}
{\vbox{\centerline{FINITE SIGMA MODELS   }
\vskip2pt
\centerline{ AND  EXACT STRING  SOLUTIONS }
\vskip2pt
\centerline { WITH  MINKOWSKI SIGNATURE METRIC}}}

\centerline{ A.A. TSEYTLIN
\footnote{$^*$}
{e-mail: aat11@amtp.cam.ac.uk
}
\footnote{$^\star$} {On leave of absence from the Department of
Theoretical Physics, P. N. Lebedev Physics Institute, Moscow 117924, Russia.}}
\bigskip
\centerline{\it  Theoretical Physics Group }
\centerline {\it  Blackett Laboratory}
\centerline{\it Imperial College}
\centerline{\it  London SW7 2BZ, United Kingdom }
\bigskip\bigskip
\baselineskip=8pt plus 1pt minus 1pt
We  consider $2d$   sigma  models with a
$D=2+N$ - dimensional  Minkowski signature target space metric  having a
covariantly constant  null Killing vector.  These models  are UV finite.
The $2+N$-dimensional target space  metric can be explicitly determined for a
class of  supersymmetric  sigma models with  $N$-dimensional `transverse'
part of the target space being homogeneous K\"ahler. The corresponding
`transverse' sub-theory  is  an $n=2$ supersymmetric  sigma model  with
the  exact $\gb$-function  coinciding with its  one-loop expression. For
example, the finite $D=4$ model has  $O(3)$ supersymmetric sigma
model as its `transverse' part.    Moreover, there exists a
non-trivial dilaton field such that the Weyl invariance conditions are also
satisfied, i.e. the resulting models correspond to  string vacua. Generic
solutions  are  represented in terms of the RG flow in   `transverse'
theory.  We  suggest a possible  application of the  constructed Weyl
invariant sigma models to quantisation of $2d$ gravity.  They  may be
interpreted as `effective actions' of the  quantum $2d$   dilaton  gravity
coupled to a  (non-conformal) $N$-dimensional `matter' theory. The conformal
factor of the $2d$ metric  and $2d$ `dilaton' are identified with  the  light
cone coordinates of the $2+N$ - dimensional sigma model.
\bigskip

 \Date{11/92}

\baselineskip=20pt plus 2pt minus 2pt
\newsec{INTRODUCTION}
One of the important problems in string theory is to classify possible
solutions of the string effective equations, i.e. string vacuum backgrounds
which may be represented in terms of Weyl invariant $2d$ sigma models (for
reviews see, e.g.,  ref.1).  Since the string equations (or `$\bgb$-
functions') are quite complicated  (already at the string tree level)
containing all terms in $\a'$  the structure of the space of solutions is
poorly understood.
Among a few classes of solutions which are explicitly
known are: (1)  flat space with linear dilaton$^{2/}$; (2) group spaces (WZW
models)$^{3/}$; (3)  `plane wave' backgrounds$^{4/}$; (4) backgrounds
corresponding to gauged WZW theories$^{5/}$; (5)  various  possible direct
products (see e.g. second paper  in ref.2).  In contrast to the first three
classes of backgrounds (which can be represented
 in a form  essentially
independent of $\a'$)  the backgrounds of the fourth type are non-trivial
functions of $\a'$ (see ref.6).  There are, of course, many other solutions of
the leading order  string equations (see, e.g., ref.7)   but their
generalisations  to all orders in $\a'$ (which
 should exist in perturbation theory) are not explicitly known. One can try
to  construct  new solutions by using  various types of duality
transformations$^{8,9,10/}$. However, since the  exact form of the  {\it sigma
model }  duality   transformations  is not explicitly known (except in  the
first two orders in $\a'$)$^{9/}$    all  discussed duality rotations of exact
string solutions solve string equations only   to the  leading order $\a'$.
  Having found an exact  string background  one is still confronting an
additional problem of identifying a conformal theory which  it
should correspond to. The solution to this problem is known only in the case of
  (gauged) WZW theories.

In  order to understand   better  gravitational applications of string
theory  (e.g.   string  backgrounds related  to
 cosmology,  black hole   physics  or  possibly   to high energy
string scattering$^{11/}$)  it
  is  important to find  new   exact solutions  which have physical
$Minkowski$ signature.   A  class of such solutions will be   described below.
In general, the solutions will be non-trivial functions of $\a'$. We shall
present a simple algorithm of their construction in terms of  the
renormalisation group flow of a non-conformal  euclidean  $2d$ theory. Namely,
the following theorem is true$^{12,13/}$ : given a non-conformal  sigma
model  with  an $N$-dimensional target space  with  euclidean signature
metric there  exists a conformal invariant sigma model in $2+N$ dimensions
with Minkowski signature metric. The $2+N$-dimensional metric depends on only
one of the two extra coordinates   (it has a covariantly constant null Killing
vector) and is expressed in terms of the ``running" coupling of the
$N$-dimensional theory (the ``transverse" part of the metric satisfies a first
order renormalisation group - type equation).    Thus  starting from  an
arbitrary $N$-dimensional euclidean  background    one can construct  a
$2+N$-dimensional  string   solution  with Minkowski signature.

 We shall  discuss the  $2d$ supersymmetric generalisation of this class
 of finite
sigma models and will show that the $2+N$-dimensional metric can be explicitly
determined in the case when     the transverse space is  homogeneous K\"ahler.
Then  the `transverse' sub-model is $n=2$ supersymmetric and the
expression for the exact $\gb$-function  of the transverse theory  is
known (it  coincides with the   one-loop result) so that the RG
equation is  easy to integrate.

Conformal invariant sigma models with a null
Killing vector  are  also of interest   in connection with the problem of
quantising $2d$ gravity.    If one starts with a $2d$  model of  gravity
coupled to a (non-conformal)  $N$-dimensional matter  theory  it  is
expected$^{14,15,16,17/}$ that  the couplings of the matter theory should
develop  a dependence on the conformal factor such that the resulting
`quantum action' is  represented by $N+1$-dimensional Weyl invariant sigma
model.  This  suggestion  suffers from the following difficulty:
 since the  Weyl invariance conditions   turn out to be $second$ order
differential equations in the $N+1$-th `time' coordinate (conformal factor)
there is an ambiguity in choosing a particular solution  which satisfies
natural  initial conditions.  This problem is (at
least partially) avoided$^{12,13/}$ if  one considers a model of  $2d$ quantum
gravity  where there is  an extra scalar  field ($2d$ `dilaton') coupled to
$2d$ curvature (see e.g. refs.18,19).
The  central observation   is that the corresponding   quantum
action can
 can be identified  with an action of a  conformal invariant
$N+2$-dimensional sigma model  with a null Killing vector. The extra scalar
field and the conformal factor play the role of the light-cone coordinates
$v$ and $u$. The theory is effectively $N+1$-dimensional  since the condition
of Killing symmetry implies that couplings are $v$-independent. As a result,
the conformal invariance equations are $first$ order differential equations
in $u$ (in fact, the standard RG equations of the `transverse' $N$-dimensional
theory) and their solution satisfying natural initial conditions is unique.

In Sec.2  we shall first show that  the sigma models with covariantly
constant null Killing vector are UV finite  in    flat $2d$ space. In
contrast to what happens, for example, in  WZW models  the
 divergences   will not cancel automatically at each order of perturbation
theory but will be absent on shell (i.e. it will be possible to redefine them
away)$^{12/}$. The mechanism of finiteness  which  operates here was  already
discussed (at the one-loop level) in ref.20.  We  shall  then study the
Weyl invariance conditions$^{21/}$  on  a  sigma model defined on a curved
2-surface and will prove (making use of the general coordinate
invariance identities for the Weyl anomaly coefficients$^{22/}$) that there
exists a dilaton field such that the sigma models with a covariantly constant
null Killing vector are  Weyl invariant$^{13/}$. That means they represent
 solutions of string  effective equations.   In contrast  to  the
 previously known string solutions with  a null Killing vector$^{4/}$    which
have     $flat$
 $N$-dimensional space   the  backgrounds we  have  found   may   have
 an arbitrary  transverse space.

A new class of finite supersymmetric  sigma models with null Killing vector
will be presented  in Sec.3.  We shall present an  explicit expression for the
$2+N$-dimensional target space metric  (with homogeneous K\"ahler transverse
subspace) which represents an exact  solution of superstring  theory and
consider some of its properties.

A relation to $2d$ quantum gravity  models will be  discussed in Sec.4.  In
particular, we shall  consider  a generalisation to the case when  the
sigma model action  contains  the
tachyonic coupling (or a  scalar potential).

\newsec{FINITENESS AND WEYL INVARIANCE OF SIGMA MODELS WITH
COVARIANTLY CONSTANT NULL KILLING VECTOR}
\subsec {Proof of Finiteness  }
  The
most general $D=N+2$ dimensional Minkowski signature metric
 admitting a covariantly constant null Killing
vector can
be represented in the form  $$ds^2 = \hg_{\mu \nu} dx^{\mu} dx^{\nu} =  -2dudv
+  \ggij (u,x) dx^i dx^j   \ \  ,
 \ \eq{1} $$ $$ \ \ \mu , \nu = 0,1, ...,  N+1 \ \ , \ \ \
  i,j = 1,...,N \ \ . $$
In fact, starting from the    null metric
$$ds^2 = \tg_{\mu \nu} dx^{\mu} dx^{\nu} =  -2dudv +  \ggij (u,x) dx^i dx^j
 + 2A_i(u,x) dx^i du + K(u,x) du^2  \
\  ,
 \ \eq{2} $$
one can eliminate $A_i$ and $K$ by a change of coordinates which preserves the
``null" structure of (2)$^{23/}$.
Thus the most general null metric is parametrized by the functions
$\ggij (u,x)$. It
is important to keep in mind, however, that (1)  considered as a generic
form of the metric is written using a specific choice of coordinates $v, x^i$.
For example, if $\ggij(u,x)$ is  flat  as a function of $x^i$ this does
not imply that  a generic  ``null" metric with a flat transverse part is just
given by $ ds^2 =-2dudv + dx^idx_i \ $: transforming  the coordinates to
make $\ggij$  equal to  $\gd_{ij}$  we will get back   the metric (2) with
non-vanishing  $A_i$ and $K$.

To establish  the UV finiteness of the  corresponding sigma model on a flat
$2d$ background one should check that  there exists  a vector $M_\mu$  such
that the $\gb$ - function  for the  target space metric $G_{\mu \nu }=\hg_{\mu
\nu}$ (1)  vanishes up to the  $M_\mu$ - reparametrisation term$^{24/}$
     $$ \gb^G_{\mu \nu} + 2 D_{( \mu} M_{\nu )} =0 \ . \eqno (3) $$
If (3) is satisfied the divergences can be   absorbed into a redefinition of
the  coordinates $x^\mu$.
 As we shall see,  (3)  is indeed satisfied for a
particular  $\ggij(x,u)$  as a function of $u$.   Using that the  non-vanishing
components of the Christoffel connection and the  curvature of $\hg$  are
  $$ {\hat \Gamma}^i_{jk} = { \Gamma}^i_{jk}\ \ ,
\ \ {\hat \Gamma}^v_{ij}=\ha \dg_{ij}
\ \ , \ \ {\hat \Gamma}^i_{ju}=\ha g^{ik}\dg_{kj}   \ \ , \ \
\ \dg_{ij}\equiv{\del \ggij \over \del u}\ \ , \  \eqno (4)  $$
 $$ {\hat R}_{ijkl} = { R}_{ijkl}\ \ , \ \
{\hat R}_{iuju}= T_{ij} \ \ , \ \ {\hat R}_{uijk} =  E_{ijk} \ \ , \eq{5} $$
 $$ T_{ij}\equiv -{1\over 2}  (\ddg_{ij}-\ha  g^{mn}\dg_{im}\dg_{nj}) \ \ , \
\ E_{ijk}= - D_{[j}\dg_{k]i}  \ \ , \eqno (6) $$
and that $\gb^G_{iv}=0,\ \gb^G_{uv}=0 \    $ (this follows from the fact that
the $\gb^G$-function is  constructed in terms of curvature tensors and
covariant derivatives)
we  can rewrite (3)  in the  `component' form
$$ \gb^g_{ij} +  2D_{( i} M_{j )} - 2{\hat \Gamma}^v_{ij}M_v =0 \ \ ,
\ \ \  $$
$$  \bgb^g_{ij} = \dg_{ij} M_v  \ \ ,
 \ \ \ \ \bgb^g_{ij}\equiv \gb^g_{ij} +  2D_{( i} M_{j )} \ \ ,
 \eqno (7) $$
$$ \gb^G_{uu}  = -2 \del_u M_u  \ \ ,  \eqno (8) $$
$$\gb^G_{iu} = - \del_i M_u - \del_u M_i + g^{jk}\dg_{ij}M_k \ \ , \eqno (9)
$$ $$ \del_i M_v + \del_v M_i =0  \ \ , \ \ \
\del_u M_v + \del_v M_u =0  \ \ .  \eqno (10) $$
Since all the components of $\gb^G_{\mu \nu}$ do not depend on $v$,
the only $v$ - dependence that is possible in $M_\mu$ is a linear $v$-term in
$M_u$. Then the general solution of (10) is given by
  $$ M_v=mu + p  \  ,  \  \ \  M_u = -mv + Q(u,x) \  , \ \ \
M_i = M_i(u,x) \  , \ \ p ,  m =\const    . \eqno (11) $$
For a given $\ggij(u,x) \ $ the components  $ \gb^G_{uu} $ and $  \gb^G_{iu} $
are some  particular  $N+1$ functions of $u$ and $ x$  so that   one can
always  satisfy  the
 equations (8) and (9)  by properly choosing  $N+1$ functions  $M_u$  and
$M_i$ (once we have solved (8), we can put (9) in the form $ \del_u M_i +
h_i^j(u,x) M_j = E_i(u,x) $ which always has a solution).

Having determined $M_u$ and $M_i$ as functionals of $\ggij$ we  are left with
the
final equation (7). It should be interpreted as an equation for $\ggij(u,x)$.
Using (11) and  introducing
 $$\tau = m^{-1} \ln  (mu+p) \ \ ,  \ \ m\not=0 \
\ ;
 \ \ \ \ \tau = p^{-1}u  \ \ , \ \ m=0 \ \ , \eqno (12)$$
(to get a Weyl invariant model one should
actually set $m=0$, see below)  we can  represent (7) in the form
$$ {d\ggij \over d \tau } = \bgb^g_{ij}  \ \ . \  \eqno (13) $$
Thus we have  proved the following  statement:  if the  metric $\ggij$
depends on $u$ in such a way that it satisfies the
standard RG equation of  the $N$
- dimensional sigma model (with some particular
reparametrisation vectors $M_i$) then the $2+N$ - dimensional  sigma model
based on (1) is UV finite to  all orders of the loop expansion.

Let us now make a number of  comments.
If $g_{ij}$ corresponds to a finite $N$ - dimensional theory, i.e. $\bgb^g_{ij}
=0$ then one should set $p=0$, i.e. a finite $2+N$ - dimensional model is found
for arbitrary dependence of $g_{ij}$ on $u$.
The above  argument for finiteness   is  simplified  in  the  ``one-coupling"
case  when   the transverse metric is proportional to a metric
$\gamma_{ij}(x)$
of a symmetric (constant curvature) space
 $$ \ggij(u,x) = f(u) \gamma_{ij}(x)    \ \ . \eqno (14) $$
The corresponding model is renormalisable for arbitrary $f(u)$. To get more
explicit formulas let us assume that the transverse space is maximally
symmetric, i.e.
$${ R}_{ijkl}(\gamma )={R\over N(N-1) }  ( \gamma_{ik}
\gamma_{jl} - \gamma_{il} \gamma_{jk}) \ . $$
 Since
$\gb^G_{iu}=0$ and scalar functions (e.g.  $ \gb^G_{uu}$) are $x$ -
independent  we set $M_i=0, \ \del_iM_u=0$ and thus  solve (7),(8),(9)
by$^{12/}$  $$ M_v=mu + p  \ \ ,  \  \ \  M_u = -mv +
Q(u) \ \ ,$$ $$  {\dot Q} = -\ha \gb^G_{uu} (u) = {1\over 4}\ga'
N(f^{-1}\ddf- \ha  f^{-2}\df^2)  + O(\ga'^3) \ \ ,
 \eqno (15) $$
 $$ M_v\df \gamma_{ij} = \gb (f) \gamma_{ij}\ \ ,  \ \ \ $$
i.e.
$${df\over d\tau } =\gb(f) \ \ , \eqno (16) $$
 $$\gb^G_{ij}= \gb^g_{ij} = \gb (f) \gij \ \ ,  \ \ \  \gb^g_{ij} = \ga' R_{ij}
+ O(\ga'^2) \ \ , $$
 $$ \gb (f) = a +
(N-1)^{-1}a^2f^{-1}  $$ $$  + {1 \over 4} (N-1)^{-2}(N+3)a^3f^{-2}  +
 O(a^4 f^{-3})   \ \  , \ \ \ \ a \equiv  \ga' N^{-1}R \ \ .  $$
 Eq.(16)  has the obvious perturbative solution (we choose $m=0$ case in
(12))
 $$
f(u) = b u + (N-1)^{-1} \ln  u + O(u^{-1})  \ \  , \ \ \ \  b\equiv p^{-1}a \
. \eqno (17) $$
 The asymptotic freedom corresponds
to $f$ (i.e. the inverse coupling of the sigma model) growing to infinity at
large $u$.  Having found $f(u)$ from (16)  one determines $Q$  from (15)
and thus solves (7)--(10).

We see that the metric of the transverse space  (and  thus the full metric
(1))  is   determined   by the  $\gb$-function of the transverse
theory.  The  explicit all order expressions for the latter are not known in
bosonic sigma models.  On the other hand, there are examples of $n=2$
supersymmetric ($n$ is the number of $2d$ supersymmetries) sigma models  with
homogeneous symmetric K\"ahler target spaces  for which the  exact
$\gb$-function   coincides  with the   one-loop expression$^{25/}$. As we shall
discuss in Sec.3, the
 metric of the corresponding  $finite$ $2+N$-dimensional $n=1$ supersymmetric
sigma models  is explicitly  given by (14) and the $first$ term in (17).

\subsec {Solution of Weyl Invariance Conditions}
 The UV finiteness of a sigma
model in flat 2-space  does not in general guarantee that the corresponding
model  on a curved $2d$ background is Weyl invariant. The Weyl invariance
conditions for the model $$ I= {1\over { 4 \pi  \ga'}} \int d^2 z \sqrt {g}
[\  G_{\mu \nu }(x) \del_a x^{\mu} \del^a x^{\nu}  +  \ga' R^{(2)} \gp (x)\ ]
\  \ \eqno {(18)}$$
(which are equivalent to the string effective equations)
have the following general structure$^{21/}$
$$ {{\bar \gb}^G}_{\mu \nu } =
\gb^G_{\mu \nu}  + 2D_{( \mu} M_{\nu )} =0 \ ,
\ \ \eqno {(19)} $$
$$ \bgb^{\phi} = \gb^\phi +   M^{\mu} \del_{\mu}\phi
=0  \ \ , \ \eqno {(20)} $$
$$  \gb^\phi = c -\ha \ga' D^2 \phi + {1\over 16}{\ga'}^2
R_{\mu \ga \gb \gg}R^{\mu \ga \gb \gg} + O(\ga'^3) \ \ ,
\ \ \ c={1\over 6 } (D - 26)\ ,  $$
where $M_\mu$ is not arbitrary but  is given by
$$M_{\mu}=\ga'\del_\mu \phi + \ha W_{\mu}\ \ . \eqno (21) $$
Here $W_\mu$ is a covariant vector constructed of $G_{\mu\nu}$ only (and
determined  by the mixing  under renormalisation  of dimension  two
composite operators$^{21/}$). To prove that sigma model based on (1) is Weyl
invariant one needs to show that there exists  a dilaton field $\phi$  such
that $M_\mu$ in (3) can be represented in the form (21).

The Weyl anomaly
coefficients  ${{\bar \gb}^G}_{\mu \nu }$  and $\bgb^{\phi}$ satisfy  $D$
differential identities which can be derived from the condition of
non-renormalisation of the trace of the energy-momentum tensor of the sigma
model$^{22/}$.  They can be  considered  to be  a consequence of the target
space
reparametrisation invariance given that
 ${{\bar \gb}^G}_{\mu \nu }$  and $\bgb^{\phi}$  are related to a
covariant effective action $S$
$$ {\d S \over \d \vp^A} = k_{AB} \bgb^B \ \ , \ \ \vp^A=(\Gmn , \ \p)\ \
,\eq{22} $$
$$ 2D_\mu {\d S \over \d \Gmn} - {\d S \over \d \p } D^\nu \p =0 \
\ . \eq{23} $$
In general, the identity  has the following structure$^{22,21/}$
$$ \del_\mu \bgb^{\phi} - {{\bar \gb}^G}_{\mu \nu }D^\nu \p
  - V_\mu^{\a \b}{{\bar \gb}^G}_{\a \b  }  =0  \ \ , \eq{24} $$
where the differential operator $V_\mu^{\a \b}  $ depends only on $\Gmn$.
To the lowest order in $\a'$ one finds$^{26,21/}$
$$ \del_\mu \bgb^{\phi} -
{{\bar \gb}^G}_{\mu \nu } D^\nu \p + \ha D^\nu ( {{\bar \gb}^G}_{\mu \nu } -
\ha \Gmn G^{\l \r}{{\bar \gb}^G}_{\lambda \r} )  + O(\a'^2) =0 \ \ . \eq{25} $$
One of the consequences of (24) is that $\bgb^{\phi}=\const$  once
(19) is satisfied. In general,  the identity (24)  implies  that  only $ \ha
D
(D+1) +1 - D$ of equations (19), (20) are independent.  It may happen, in
particular, that  if  the ``transverse" subset of $\ha (D-2)(D-1) $  equations
in (19) and  the dilaton equation (20) are solved, the remaining $D$
equations
(19) are satisfied automatically.

Let us look for solutions of (19),(20) which have the form$^{12,13/}$
$$ \Gmn = \hg_{\mu \nu} (u,x) \ \ , \ \ \ \p = \p (v,u,x) \ \ , \ \ x^\mu=
(v, u, x^i) \ \ , \eq{26} $$
where $\hg_{\mu \nu}$ is given by (1).
Since $\gb^G_{\mu \nu} \ , W_\m $  and hence
 $\gb^G_{\mu \nu}{}'=\gb^G_{\mu \nu} +   D_{( \mu} W_{\nu )} $  in (19) are
covariant functions of the curvature and its derivatives   and since the
metric has a Killing  vector it is easy to see that   the  $( \m v)$  component
of $\gb^G_{\mu \nu}{}'$  is identically zero. Then (19) gives the following
constraint on the dilaton:    $ \del_\m \del_v \p =0\  , \ $ i.e. $$ \p = p
v +  \p (u,x) \ \ , \   \ \ \  \ p=\const  \ \ . \eq{27} $$ Here $p$ is an
arbitrary integration constant
 and $\p (u,x)$ is to be determined. From now on all the functions  will depend
only on $u$ and $x^i$. Using (4), (27)  we can represent the non-trivial
components of (19)  as follows (we shall put $\a'=1$)
$$ \bgb^g_{ij}  - p \dg_{ij} =0 \ \
,  \ \ \eqno {(28)} $$
$$
 \bgb^g_{ij} \equiv \gb^G_{ij} +   D_{(i} W_{j )} + 2  D_{i} D_{j} \phi
\ \ , $$
$$ \gb^G_{iu} + \ha \del_i W_u + \ha \dot W_i - \dg_{ij}W^j + 2 \del_i\dpp
-\dg_{ij} D^j \p =0 \ \ ,  \eq{29} $$
$$ \gb^G_{uu} + \dot W_u  + 2 \ddp
=0 \ \ .  \eq{30} $$
Equation (20)  takes the form
 $$ \bgb^{\phi} = c - \gamma \p + (\del_\mu \p)^2  + \ha W^\mu
\del_\mu \p + \o  $$
$$ ={1\over 3 } + {\bgb^{\phi}}{}' + \ha  p M^{ij}
\dg_{ij}
 - \ha p W_u - 2p \dpp
 =0  \ \ , \ \eqno {(31)} $$
$$ {\bgb^{\phi}}{}' \equiv  c'  - \g' \p + (\del_i \p)^2  + \ha W^i \del_i
\p  +\o \ \ , \ \ c'={ 1\over 6}  (N-26) \ \ \ ,
\eq{32} $$ where $\g'$ is the `anomalous dimension' differential  operator,
$\o$ is a
 covariant function of $G_{\mu\nu}$ only and the $M^{ij}$-term ($M^{ij} = \ha
g^{ij} + ...$) in (31) originates from  the linear in $\p$ term $ - \g \p= -
\g' \p - M^{ij} D_i D_j \p + O(D^3 \p) $  (see ref.13 for details). Being
scalar functions of the curvature $\g'$,  $\o$ and
hence $ \bgb^{\phi}{}'$ do not depend on the  derivatives of the metric over
$u$.
  The functions $\bgb^g_{ij}$
 and ${\bgb^{\phi}}{}'$   can be interpreted as the Weyl anomaly
coefficients of the ``transverse" theory defined by $g_{ij}(u,x)$ and $\p
(u,x)$ at fixed $u$ (${1\over 3}$ in (31)  corresponds to the  central charge
contribution of the two light-cone  coordinates).

 Let us first consider the case of $non-vanishing$  $p$.
Then (28) is a first order differential  equation for
$g_{ij}(u,x)$ which always has a solution. Eliminating the derivatives of
$g_{ij}$  over $u$ from (31) using (28) we  find  a similar first order
equation for $\p (u,x)$. Eqs. (28),(31) can be interpreted as renormalisation
group equations of the ``transverse" theory with $u$ playing the role of the RG
``time"$^{12/}$.

  Still the  is a  question   whether the solutions of (28) and (31)
satisfy also (29) and (30). It  is answered  positively$^{13/}$ using the
identity  (24).  Substituting  $ \bgb^G_{ij}=0\ , \ \  \bgb^\phi =0$ and the
expression  (27) for the dilaton into (24)  one finds$^{13/}$
 $$  p \bgb^G_{i u} =0\ \ , \ \ \ \  p \bgb^G_{u u} - \bgb^G_{ iu}
D^i \p - 2 V_u^{ju}{{\bar \gb}^G}_{ju} =0\ \ . \eq{33} $$
That means  that once (28) and (31) are satisfied for non-zero $p$
the remaining equations (29) and (30) are satisfied as well.
 The conclusion is that  given  some  initial data $( \gij (x) , \ \p (x)
)$ at $u=0$ there exists  a  $u$ - dependent solution $ ( \gij (u,x) , \ \p
(u,x) ) $ of the Weyl invariance conditions (19)--(21).

In the particular case when the transverse  space is  symmetric  (i.e.
its  metric is given by (14)) the symmetry requires that $W_i=0,\
\bgb^G_{i u}=0$ and that $\phi$ is $x^i$-independent,
$$ \p = pv +  \p (u) \ \ . $$
The  functions  which  enter the   equations  for  $f(u)$ and $\p (u)$
 are
$$\gb^G_{ij}= \gb (f) \gamma_{ij} \ , \ \  \ \gb^G_{uu}= \gb^G_{uu} (f) \ , \
\  \ \  W_u= W_u (f) \ , $$ $$ \ \ {\bgb^{\phi}}{}'  =  c'
  +\o  (f) \ , \ \ \ \
 M^{ij}=  \ha f^{-1}[1+M(f) ]\gamma^{ij} \ \ .
$$  Since eq.(30) is a consequence of (28) and (31)  $\gb^G_{uu}$ is not
independent and we are left with  the following two equations for $f(u)$ and
$\p (u)$ (cf.(15),(16))
$$ p \dot f = \gb (f) \ \ , \ \ \ \  p \dot \p = \ha c + J(f)  \ , \ \  \eq{34}
$$ $$   J= \ha \o (f) + {1 \over 8}     N
 [1 + M(f) ] f^{-1} \gb (f)  - \fourth p W_u  \ , \ \ \
c = {1\over 6 } + c'={1\over 6 } (N - 24)\ .
  $$
As a result, the `scale factor' of the metric $f(u)$ runs according to the
standard (``flat  space") RG equation while the dilaton depends  on $u$
 in such a way as  make  the total central charge vanish.
It is possible to show$^{12/}$ that if (28),(30) are satisfied
 the central charge of this   model $\bgb^{\phi}$ is  equal
to that of the free  $2+N$ - dimensional theory plus the contribution of the
linear terms in the dilaton. In fact, since $\bgb^\phi$ is constant
on a solution of (28),(30)
 it   can be computed at any value of
$u$, e.g. $u=\infty$. Given that all higher loop contributions should vanish in
the weak coupling limit of large $u$ (we are assuming that the transverse
sigma model is asymptotically free) it is sufficient to  compute $\bgb^{\phi}$
in the leading order  approximation. Representing
the dilaton in the form
$$ \p = pv + qu  +  \bp (u) \ \ , \eq{35} $$
where $\bp$  stands for  contributions which are due to  sigma model
interactions (which depend on the coupling $f$, i.e. $\bp (u)= F(f(u))$) we
find  that  the `free theory' and `interaction' contributions cancel
separately,   giving
$$ \bgb^\phi = c - 2pq  =0 \ \ ,  \ \ \ \
 p {\dot {\bp}} = J(f(u))  \ . \eq{36}  $$
Thus one can satisfy the zero total  central charge condition for arbitrary $N$
by a proper choice of the constants $p$ and $q$.

If the ``initial"
transverse theory  is  generic, i.e. if $\bgb^g_{ij}$ in  (28)  is
non-vanishing at $u=0$ then  the  solution exists only for a non-zero $p$.
If, however, the initial theory is Weyl invariant, i.e.
 $$\bgb^g_{ij}(u=0) =0 \ \ \ , \ \  \ \ {\bgb^\p }{}' (u=0) =
c'' = \const \ \ , \  $$
  there are two possibilities.    For $p\not=0$   the
simplest solution   of  (19),(20) is the `direct
product'  one    represented by  the fixed point
 of the RG equations (28),(31)
  $ \gij (u,x)=\gij (x) \ ,
   \ \p (u,x) = {1\over 2p}( {1\over 3 } + c'')
  u + \p (x) . $
  When
the ``transverse" theory $ (\gij (u,x)  ,  \ \p (u,x)) $ is Weyl invariant at
$u=0$  $and$  $p=0$   eqs.(28),(31) imply that
  the  initial Weyl invariance conditions
(34) are satisfied  also  for  all  other values of $u$. Therefore  a
solution with  (34),(27) and $p=0$  may exist only if the transverse theory
is conformal for all $u$.  One can also prove  the  converse$^{13/}$: to get a
non-trivial solution   with a flat
 $\gij (u,x)$ (more generally, with a conformal transverse theory)  one should
set $p=0$.  Then  (assuming $\p = \p (u)
$) eqs.(28),(31)   are  satisfied automatically  but since $p=0$ the
identities (33)  {\it no longer imply}  that  (29),(30) are also
satisfied.

 Since  (28) holds
identically  it does not  give an equation for $\gij (u,x)$. The same is
true for (31): it does not contain  terms with $u$ - derivatives  and being
a constant (as a consequence of (19),(24))  is satisfied  for  all $u$
if it  is  satisfied    for $u=0$, i.e. if $ \third + c' =0$.
 Instead of $N+1$ identities  in (33)     for $p=0$ we are left with
just one. As a result,  we get $N$  independent equations (29),(30) ((33) gives
a relation  between   components of (29))
on  $\ha N(N+1) + 1 $ functions
$\gij (u,x)\ , \ \p (u,x)$.
Their particular solutions in the case when the transverse metric is flat
 (and correspondence with the  `plane-wave' solutions found
previously$^{2/}$)  were studied in detail in ref.13. In that case it is useful
to change   coordinates, trading the functions $\gij (u,x)$ corresponding
to a flat transverse metric  for $A_i$ and $K$ in (2) , i.e.   transforming
the metric (1) into the form
 (2) where $\gij (u,x)$ has its  canonical $\gd_{ij}$ form.

 The above
 discussion  can be generalised to the case of non-vanishing antisymmetric
tensor coupling$^{13/}$. Namely, there  exist  similar solutions of the Weyl
invariance conditions with the metric (1), dilaton (27) and the
$v$-independent antisymmetric tensor $\B_{\mu\nu}: $
$\ \B_{ij} = B_{ij}(u,x)\  , \  \  \B_{iu} = B_i (u,x) \ , \ \ \B_{\mu v
} =0 .  $

\newsec{  NEW CLASS OF FINITE SUPERSYMMETRIC  SIGMA  MODELS  WITH MINKOWSKI
SIGNATURE TARGET SPACE}
 In Sec.2 we have shown  that   it is possible to
construct conformal invariant Minkowski signature models in $2+N$ dimensions
from  non-conformal Euclidean
  models in $N$ dimensions.  Since the metric and the
dilaton of the $2+N$-dimensional theory are essentially the `running'
couplings of the transverse theory  their dependence on $u$  is determined by
the $\gb$-functions of the transverse theory.
The structure of the $\gb$-functions is usually  simpler in
supersymmetric theories  so it is of interest to generalise the above
construction to the  supersymmetric case.  In particular, we would like to
make use of the known fact that there are examples of
supersymmetric  sigma models  with
homogeneous symmetric K\"ahler target spaces  for which the  exact
$\gb$-function   coincides  with the   one-loop expression, i.e. is
explicitly  calculable$^{25/}$.

The two dimensional ($n=1$) supersymmetric  sigma
model can be constructed for an arbitrary metric $G_{\mu\nu}$ of a
$D$-dimensional  target space. Its  superfield  action is given by$^{27/}$
$$ I= {1\over { 4 \pi  \ga'}} \int d^2 z d^2 \t  \  G_{\mu \nu }(X)
\D X^{\mu} { {\bar \D } } X^{\nu} \  \ , \ \eqno
{(37)} $$
where
$$ X^\mu = x^\mu + \bt \psi^\mu + \half \bt \t F^\mu\ \ , \ \ \
\D= {\del \over \del \bt } + \bt \gamma^a \del_a \ \ . $$
The component form  of the action is
$$ I= {1\over { 4 \pi  \ga'}} \int d^2 z \  [  G_{\mu \nu }(x)
\del_a x^{\mu} { \del^a } x^{\nu}  +
G_{\mu \nu }(x){\bar \psi}^\mu \gamma^a D_a \psi^\nu +
{1\over 6} R_{\mu\nu\lambda \rho}{\bar \psi}^\mu{ \psi}^\lambda
{\bar \psi}^\nu{\psi}^\rho ]  \ . \ \eqno
{(38)}$$
For the metric with the null Killing  vector  (1)   we can represent (37) in
terms of the real superfields $U,\ V$ and $X^i$
$$ I= {1\over { 4 \pi  \ga'}} \int d^2 z d^2 \t \  [ -2 \D U { \bar \D } V +
g_{ij}(U, X) \D X^{i} { \bar \D } X^{j} ] \  \ . \ \eqno
{(39)}$$
The component form of (39) can be found  either directly from (39) or by
substituting the expressions  (1),(4),(5),(6) into (38).

Eqs.(3)--(17) have a straightforward generalisation to the supersymmetric case.
In particular, the solution $g_{ij}(u,x)$ of the condition of finiteness (13)
is determined by the $\gb$-function of the `transverse'  part of
(39), i.e. of the supersymmetric model with the metric  $g_{ij}(u, x)$  for
constant $u$.  As is well known$^{28/}$, if  the transverse space is K\"ahler
the $N$-dimensional model is $n=2$ supersymmetric.  If
it is also a   compact symmetric homogeneous space (e.g. $S^2= SO(3)/SO(2)$
or $CP^m$)  then   it is very plausible that its $\gb$-function is exactly
calculable  and is  given  by the one-loop expression$^{25/}$. This   was
actually proved in ref.25 for the following classes of  K\"ahler manifolds:
  $$
M_1= SO(m+2)/SO(m)\times SO(2) \ ,  \  \ \ N=2m \ ; $$ $$
\ \ M_2= SU(m+k)/SU(m)\times SU(k)\times U(1)\  , \ \  \ N = 2mk\ ;  $$ $$  \
\  M_3=Sp(m)/SU(m)\times U(1)\ , \ \ \ N=m^2+m\  ;  $$ $$  \ \
M_4=SO(2m)/SU(m)\times SO(2)\ , \ \ \  N= m^2-m\ . $$ In that case the
transverse part of the metric (14), the $\gb$-function   (16)  and the
solution of (13) are  given simply  by
 $$ \ggij(u,x) = f(u) \gamma_{ij}(x)    \ \ , \ \ \ \gb (f) = a  \ \ , \ \ \
\  \  f(u) = b u  \ , \ \ \ b=p^{-1} a \ . \eqno (40) $$
The constant  $a > 0$  is  determined by the geometry of
the  transverse space$^{25/}$ ($a_1(m=1) = 2 \  ; \  a_1( \ m\geq 2)= m \  ;\
a_2= m +k \ ; \ a_3= m +1\ ; \ a_4= m -1 $).
The
constant $b$ is  arbitrary and  can be   absorbed  into  a redefinition
of the  coordinates $u$ and $v$. Then the final expression for the  Minkowski
signature
 metric of  the  finite  $2+N$-dimensional   supersymmetric
sigma model is
 $$ds^2  =  -2dudv
+  u \gamma_{ij }(x) dx^i dx^j   \ \   \ \eq{41} $$
(we have assumed $u>0$).
Note that while the transverse model (with fixed constant
$u$)  is $n=2$ supersymmetric the  full $2+N$-dimensional model
apparently  has  only $n=1$ supersymmetry.   The  non-zero components of the
curvature of the metric (41) can be  found from  (5),(6)
 $$ {\hat R}^i_{jkl} = { R}^i_{jkl}(\gamma )
 \ \ ,  \ \ \    \
{\hat R}_{iuju}= {b\over 4u}  \gamma_{ij} \ \ . \eq{42} $$
 All  curvature invariants are singular  at $u=0$. It is still possible that
this singularity is harmless in string theory (cf. ref.29).

The  simplest non-trivial example of the finite models we have constructed
corresponds to the case when the transverse theory is represented by the
$O(3)$ supersymmetric sigma model$^{30/}$. The resulting metric (41)
is that of $four$ ($2+N=4$) dimensional space  with the transverse part
being  proportional to the metric  on $S^2$,
  $$ds^2  =  -2dudv
+  u (d\t^2 + \sin^2 \t d\varphi^2 )   \ \  .  \ \eq{43} $$
This metric is conformal to the standard metric on
the product of  the two-dimensional Minkowski space and two-sphere.
The corresponding geodesic equations can be easily integrated  with the
conclusion that the  part of  space  with $u>0$
is not geodesically complete (replacing the factor $u$ by the modulus $|u|$
 apparently introduces additional singularities  at $u=0$).

To find out whether the constructed  finite  supersymmetric models  can be
identified with  the exact solutions of the superstring  effective equations
we need to check that these sigma models  correspond to  Weyl invariant
theories  on a curved $2d$ background. It is straightforward to add to (37)
the dilaton coupling term  $ \int d^2 z d^2 \t  E^{-1} R^{(2)}  \p (X) $
($E^{-1}$ is the determinant of the $n=1$ supervielbein)  and to generalise
the  expressions for  the Weyl invariance conditions (19)--(21) and the
identity (24) to the case of  $n=1$ supersymmetric sigma models$^{31/}$.
 Then the  argument in Sec.2.2  can be repeated  to prove
that for an arbitrary ``initial" ($u=0$) transverse euclidean $n=1$
supersymmetric model  there exist  such  metric $g_{ij}(u,x)$ and dilaton
$\p (u,x)$  that the corresponding $n=1$  supersymmetric model  with metric
(1) is Weyl invariant, i.e. represent a string vacuum.

Let us now specialize to the case when the transverse metric is  symmetric
K\"ahler.  Then we can apply the  discussion of the symmetric transverse space
case  in  Sec.2.1.  We conclude that   the  Weyl
invariance conditions are  again given by  equations (34),(36).  The  equation
on $f$ is the same  RG equation as  in the finiteness condition so its solution
is  represented  by (40),(41).   Since the transverse model is $n=2$
supersymmetric we can make use of the  result$^{31/}$ that the dilaton
coupling  is not renormalised  in the $n=2$ supersymmetric
case (in the minimal subtraction scheme). That means that $M$ and $\o$ in  $J$
in (34)  should vanish.  As a result, the dilaton $\p$ is given by
(cf.(35),(36)) $$ \p (v,u)  = pv + qu  +  \bp (u) \  ,  \ \ \ \
\bgb^\phi = c - 2pq  =0 \ ,  \ \ \  c=\fourth (N-8)\ \ , \ \eq{44} $$
 $${\dot \bp  } =  I(f(u))\ ,  \ \ \  I= p^{-1}J=   {N\over 8p}f^{-1}
  \gb (f)  - \fourth  W_u= {N\over 8u}   - \fourth  W_u \ ,  \ \
\ f= bu \ ,  \eq{45} $$ where we have used that in the superstring  theory $
c= \fourth (D-10)$.
  Note that differentiating  the equation for $\bp $ in
(45) and comparing with (30)   gives
$$  \dot W_u = - {N\over 2 p} {d  \over du} (f^{-1} \gb )- 2 \gb^G_{uu} =
\ha N u^{-2} - 2 [ \fourth  N u^{-2} + O(\a'^3u^{-3})]=O(\a'^3u^{-3})
\eq{46} $$
(the one-loop term in $\gb^G_{uu}$, i.e. $\a'R_{uu}$,   is given
by (42); see also (15)).  Higher loop corrections to $W_u$ and to
$\gb^G_{uu}$ are thus directly related. It is easy to see that there
is no two-loop term  in $\gb^G_{uu}$ in the case of symmetric transverse
space; in the bosonic case both $\gb^G_{uu}$  and $W_u$  are
non-vanishing in the tree-loop approximation$^{12/}$.

 It is possible that  $W_u$
is actually   vanishing   in the present model. Though  the results of ref.32
(a comparison of the  perturbative expression for the  $\gb^G$-function with
superstring effective equations) imply that $W_\mu$
contains a non-zero four loop term  in a general $n=1$ supersymmetric
model,  $W_\mu$  does vanish in   $n=2$ supersymmetric
models$^{30/}$. If  $W_u=0$  then  the exact expression for the
dilaton is (see (44),(45)) $$\p (v,u)  =  \p_0 +  pv + qu  +  {1 \over 8 } N
\ln u \ . \eq{ 47 } $$
The  resulting backgrounds (41),(44) or (47) thus represent
exact solutions of superstring effective equations with non-trivial dilaton.
 Note that the string coupling $\exp \phi$ is
$$ e^{\phi} = A u^{N/8} e^{(qu +   pv)}\ \ , \  \ \ \ \ A=e^{\p_0} \ . \ \eq{
48} $$ It  goes to zero  in the  strong coupling region $u \ra 0$ of the
transverse sigma model  , i.e. is $small$ near the singularity $u=0$.  If
$N<8$  the constant $q$ in (44),(47) is negative (we are assuming $u>0, \ v>0
,\  p>0$) so  that the string coupling is also vanishing in the small coupling
region  $u \ra \infty$.
In  the case   of
the critical dimension $D=10$ (or $N=8$)   $q$  must
vanish. Then  the string coupling is inverse proportional to the sigma model
coupling $f^{-1}$,
$$e^{\phi} = A'  f  e^{  pv}\ .$$

 \newsec {APPLICATION TO $2D$ QUANTUM GRAVITY  }
As is well known,  the classical gravitational action in $d=2$ is trivial
 before one accounts for the (non-local) quantum anomaly
term. Introducing an extra scalar field (``2d dilaton") coupled  to the
scalar curvature one  obtains a non-trivial theory (though still with no
propagating degrees of freedom). This theory   seems simpler and better
defined   as a starting point for a (perturbative) quantisation. By
redefining the fields one can represent  the general action in the
form$^{18,19/}$ $$S=- \ha \int d^2x \sqrt{ g}\big[
\del^\mu\varphi
\del_\mu\varphi+ q\varphi  R + V(\varphi)
\big] \ .\eqno(49)$$
For example, the
metric -- dilaton action
 which generates the $\gs $-model Weyl anomaly
coefficients in the case of $D=2$ target space
and which has a classical ``black hole" solution$^{33/}$
$$S=- \ha\int d^2x \sqrt {g} e^{-2\phi}\big[ R+ 4(\del \phi )^2
+c\big] \ ,\eqno(50)$$
 can be represented as (49) with $V=  c \exp {(\vp /q)}\ $.
By a further redefinition it  can be put into the form
$$S=- \ha \int d^2 x \sqrt{\hat g}
\big( \hat R v +  c \big)\  \ , \ \  \
\hat g_{\mu \nu} =
 v g_{\mu \nu } \ , \ \  v =  e^{-2\phi}\ .
 \eq{51} $$
Let us now switch to `world sheet' notation and consider the
 metric-scalar  (${\hgg}, v $) gravitational  theory  coupled to some extra $N$
``matter" scalar fields  which is described by the  sigma model
$$ I_0 = {1\over { 4 \pi  }} \int d^2 z \sqrt {\hgg} [\ p v \hR^{(2)} +
\gij (x) \del_a x^{i} \del^a x^{j} + T (x) \ ]\ \ , \eq{52} $$
 In the conformal gauge
 $$ \hgg_{ab} = e^{-2u/p} \g_{ab} \ \  $$
 (52)  takes the form
$$ I_0 = {1\over { 4 \pi  }} \int d^2 z \sqrt {\g} [\
- 2 \del_a v \del^a u  +  \gij (x) \del_a x^{i} \del^a x^{j}
 + p v R^{(2)} + T (x)\ e^{-2u/p} \ ] \ \ . \eq{53} $$
This model is renormalisable on a flat background with $\gij$  `running'
with a cutoff. Once all the fields are quantised  one may  expect
that  the `effective action' will be represented by a general sigma model in
$2+N$ dimensions  $x^\mu= (u,v,x^i)$. The model should be
Weyl  invariant with respect to the background metric $\g_{ab}$ since the
$2d$ metric itself  is an integration variable$^{14,15,16/}$.
We are implicitly assuming that
   the theory can be regularised in a  way covariant
with respect to the  original metric ${\hgg}$ so that  all the elements of the
theory - the action, the measure and the regularisation depend only on the
full  ${\hgg}$ (and that we are in the phase
 where $2d$ metric has zero expectation value). To determine the
`effective action'  we need to find a
solution of the Weyl invariance conditions for the metric, dilaton and tachyon
couplings  of the  $2+N$-dimensional theory  such that at  the classical limit
they reduce to the couplings in (53).  It seems   natural to impose an
additional assumption that the dependence of the couplings on $v$   in  the
`effective action' should remain as simple  as in (53), i.e. the target space
metric  and the tachyon should  be  $v$-independent (the metric will have
a Killing vector) while   the dilaton will be at most linear in $v$.
It is precisely such solutions of the metric and dilaton Weyl invariance
conditions (19),(20) that we have studied in Sec.2 (let us first ignore the
tachyon coupling  term). We have found that  the
action  $$I= {1\over { 4 \pi  }} \int d^2 z \sqrt {\g} \ [\
- 2 \del_a v \del^a u  +  \gij (u,x) \del_a x^{i} \del^a x^{j}
  + (pv + \p (u,x))   R^{(2)} \ ] \  \eq{54}   $$
defines a Weyl invariant quantum theory if  the metric $\gij $ and dilaton
$\p $ depend  on  $u$ according to the first order RG equations (28),(31).
The  result that $\gij $ starts running with $u$ according to the RG equation
$ \dg_{ij} \sim  R_{ij} +...$ is very natural given that
$u(z)$ is  proportional to the conformal factor of the $2d$ metric (which
should be coupled to a covariant cutoff).
At the same time,   one would  also expect  to find
the conformal anomaly term $\sim K(u,x) (\del u )^2 $
  but it is
missing in (54).  Note, however, that such term can be generated by  a
redefinition of the field $v$.  As discussed in ref.13,  there is, in fact,
an   $equivalent$ solution of the conformal invariance conditions (19)--(21)
with  $\p (u,x) =0$ but  with the metric (1) containing the additional term
$K(u,x)du^2$   (cf.(2)).  The  difference  between  the theory
(52)  and  the standard $2d$ gravity coupled to a sigma model (where
both  the anomaly term  $ K(u,x) \del_a u \del^a u $ and $ \p (u,x)
R^{(2)}$  should  appear in the quantum action$^{16/}$) is due to the presence
of the  extra  scalar field $v$.

Let us now   study the solutions of the Weyl invariance condition for
the  tachyon coupling$^{34,21/}$ (cf.(19)--(21))
$${\bar \beta}^T=-\g T + ( \a' \del^\mu \phi + \ha W^\mu ) \del_\mu T
-2T +  b(T) $$
$$ = - \ha \a' D^2 T + \a' \del^\mu \phi \del_\mu T  -2T + O(\a'^3) + b(T)
=0 \ \ . \eq{55} $$
 $\g$ is the same  differential operator  which appeared in (31). $b(T)$
represents  ``non-perturbative"
corrections which are of higher order in $T$.
If  there were no $v$ coordinate  so that  the metric of the $1+N$-dimensional
space  and the dilaton were given by  $ds^2 =  K du^2 + ds^2_N $  and $ \p =
Ku + ...$  then (55) would  reduce to  a second order  equation in $u$
$^{17/}$,  $ \  - \ha K^{-1} {\ddot T}  + {\dot T } + ... =0 $,
which would reproduce  the standard RG equation only in the ``semiclassical"
limit of large anomaly coefficient $K$.
 On the other hand, if  the metric $G_{\m \n}$ is given by (1) and
the dilaton is linear in $v$ (27) then for $v$-independent  tachyon $T=T(u,x)$
eq.(55) takes the form similar to (28),(31), i.e.  it becomes  a {\it first
order} RG-type equation(cf. ref.17)
$$ p{\dot T} = {\bar \b}^T{}' \ \ \ .
\eq{56} $$ ${\bar \b}^T{}'$ (containing only derivatives over $x^i$) denotes
the Weyl anomaly coefficient of the `transverse' theory with the coupling
$T(u,x)$ and $u=const$ playing the role of the RG ``time".
The simplest example of a solution of (55),(56)
is found if $T=T(u)$. Let us   first ignore  the
 ``non-perturbative" term $b(T)$. Then (cf.(53))
$$ p {\dot T}=- 2 T \ \ \ , \ \ \ \  T = T_0 e^{ -2 u/ p } \ \ . \eq{57} $$
Equivalent solutions in the context of $2d$ gravity model were discussed in
ref.19.  Now it is possible show that $T$ in (57) solves  the
full eq.(56) (with all higher order terms included), i.e. that
there are no non-perturbative divergences
in the model
$$I = {1\over { 4 \pi  }} \int d^2 z \sqrt {\g}\  [\
- 2 \del_a v \del^a u  +
pv R^{(2)} + T (u) \ ] \ \ . \eq{58} $$
In fact,  $v$ plays the role of a Lagrange multiplier
  which makes $u$ effectively non-propagating so that there are
no quantum corrections in the theory (see also ref.35). Then the condition of
conformal invariance is equivalent to
 the classical conformal invariance relation (57). To reconcile this
conclusion with the expected presence of $O(T^2)$  and $O(\del T \del T )$
terms in ${\bar \b}^T$, ${\bar \b}^\phi$ and ${\bar \b}^G$ one is to note
that   a derivation of such terms (or a proof of correspondence with $O(T^3)$
terms  in
the effective action) presumes an analytic continuation in momenta and is not,
strictly speaking, valid in the case when $T$ depends  just on one variable
(the question of non-perturbative terms in the $\b$-functions should be
addressed separately for each $2d$ theory corresponding to a  particular scalar
potential $T$, see ref.34).

In conclusion, we have suggested a  connection between the
conformal invariant  $2+N$ - dimensional sigma models and the  $2d$
scalar quantum gravity coupled to  non-conformal `transverse' $N$ -
dimensional  sigma models. The conformal factor of the $2d$ metric  is
identified not with  time but with the light cone coordinate $u$; this  makes
the corresponding Weyl invariance conditions first order in $u$.
Given that  the target space  metric corresponding to $2d$   gravity plus
scalar matter  models  has  natural  Minkowski signature$^{18/}$ it seems
important to try  to clarify further the  connection between   the `Minkowski'
conformal theories and $2d$ quantum gravity.

\bigskip\bigskip


I have benefitted from  conversations  with M.
Shifman,  K. Stelle and Yu. Obukhov.
I would like to acknowledge  a support of SERC.
\bigskip \bigskip
\bigskip
\centerline {REFERENCES}
  \item {[1]} C. Callan and L. Thorlacius, in: Proc. of 1988 TASSI School,
p.795 (Providence, 1989); \ \
A.A. Tseytlin, Int. J. Mod. Phys. A4(1989)1257.
\item {[2]}  R. Myers, \pl B199(1987)371;
I. Antoniadis, C. Bachas, J. Ellis and D. Nanopoulos,
\pl B211(1988)393.
\item {[3]} E. Witten, Commun. Math. Phys. 92(1984)455.
\item {[4]}  R. Guven, \pl B191(1987)275;
D. Amati and C. Klimcik, \pl B219(1989)443;
G. Horowitz and A.R. Steif, \prl 64(1990)260;
\pr D42(1990)1950;
G. Horowitz, in:  Proceedings
of  Strings '90,
College Station, Texas, March 1990 (World Scientific,1991); R.E. Rudd, \np
B352(1991)489;
C.Duval, G.W. Gibbons and P.A. Horv\'athy, \pr D43(1991)3907.
\item {[5]}
I. Bars and D. Nemeschansky, \np B348(1991)89;
 E. Witten, \pr D44 (1991) 314; I. Bars, in: Proc. of XX-th Int. Conf.
on Diff. Geom. Methods in Physics,
eds. S. Catto and A. Rocha (World Scientific, 1992);
 J.H. Horne and G.T. Horowitz, \np B368 (1992) 444;
M. Crescimanno, \mpl A7(1992)48;
 I. Bars and K. Sfetsos, \mpl A7(1992)1091;
P. Ginsparg and F. Quevedo, Los Alamos preprint LA-UR-92-640, 1992.
\item {[6]} R. Dijgraaf, E. Verlinde and H. Verlinde, \np B371(1992)269;
A.A. Tseytlin, \pl B268(1991)175;
I. Jack, D.R.T. Jones and  J. Panvel, preprint LTH-277(1992);
I. Bars and K. Sfetsos, preprint USC-92/HEP-B2, 1992.
\item {[7]} M. Mueller, \np B337(1990)37;
D. Garfinkle, G. Horowitz and A. Strominger, \pr D43(1991)3140.
\item {[8]}G. Veneziano, \pl B265(1991)287.
\item {[9]} A.A. Tseytlin, \mpl A6(1991)1721; in :  Proc. of
 Strings and Symmetries, NATO Advanced Research Workshop, Stony Brook, May
20-25, 1991, p.173 (World Scientific, 1992).
\item {[10]} K.A. Meissner and G. Veneziano,
 \pl B267 (1991) 33; \mpl A6(1991)339;
A. Sen, \pl B271(1991)295;
M. Gasperini, J. Maharana and G. Veneziano, \pl B272(1991)277;
S.F. Hassan and A. Sen, \np B375(1992)103; preprint TIFR-TH-92-61.
\item {[11]} D. Amati,  M. Ciafaloni and  G. Veneziano, \ijmp A3(1988)1615.
\item {[12]} A.A. Tseytlin, \pl B288(1992)279.
\item {[13]} A.A. Tseytlin,  preprint DAMTP-92-49, \np (1992).
\item {[14]} F. David, \mpl A3(1988)1651; J. Distler and H. Kawai,
\np B321(1989)509; S.R. Das, S. Naik and S.R. Wadia, \mpl
A4(1989)1033;
J. Polchinski, \np B324(1989)123.
\item {[15]} T. Banks and J. Lykken, \np B331(1990)173.
\item {[16]} A.A. Tseytlin, \ijmp A5(1990)1833.
\item {[17]} S. Das, A. Dhar and S. Wadia, \mpl A5(1990)799;
 A. Cooper, L. Susskind and L. Thorlacius, \np B363(1991)132;
A. Polyakov, Princeton  preprint PUPT-1289 (1991).
\item {[18]}
C. Teitelboim, \pl B126(1983)41; R. Jackiw, in: Quantum Theory of
Gravity, ed. S.Christensen (Adam Hilger, Bristol 1984);
A.H. Chamseddine, \pl B256(1991)2930; \np B368(1992)98;
T. Banks and M. O'Loughlin, \np B362(1991)649.
\item{[19]}C.G. Callan, S.B. Giddings, J.A. Harvey and
A. Strominger, Phys.  Rev.
D45 (1992) 1005;  J. Russo and A.A. Tseytlin,\np
B382(1992)259;
H. Verlinde, preprint PUPT-1303;
A. Strominger, preprint UCSBTH-92-18;
T.T. Burwick and A.H. Chamseddine, \np B384(1992)411;
S.P. de Alwis,  preprint COLO-HEP-280(1992);
A. Bilal and C. Callan, preprint PUPT-1320;
\item { }S.B. Giddings and A. Strominger, preprint UCSBTH-92-28.
\item {[20]} B. de Wit, M.T. Grisaru, E. Rabinovici and H. Nicolai,
 \pl B286(1992)78.
\item {[21]} A.A. Tseytlin, \pl B178(1986)34; \np B294(1987)383.
\item {[22]}  G. Curci and G. Paffuti, \np B268(1987)399.
\item {[23]} H.W. Brinkmann, Math. Ann. 94(1925)119.
\item {[24]} D. Friedan, \prl 51(1980)334 ; Ann. Phys. 163(1985)318.
\item {[25]} V. Novikov, M. Shifman, A. Vainshtein and V. Zakharov, \pl
B139(1984)389; Phys. Rep. 116(1984)103;
A. Morozov, A. Perelomov and M. Shifman, \np B248(1984)279.
\item {[26]} C.G. Callan, D. Friedan, E. Martinec and M.J. Perry, \np
B262(1985)593.
\item {[27]}  L. Alvarez-Gaum\'e and    D. Freedman, \cmp 80(1981)443;
L. Alvarez-Gaum\'e ,   D. Freedman and S. Mukhi, \ap 134(1981)85.
\item {[28]} B. Zumino, \pl B87(1979)203;
L. Alvarez-Gaum\'e and   D. Freedman, \pr D22(1980)846
.\item {[29]} S. Cecotti and C. Vafa, \prl 68(1992)903.
\item {[30]} P. Di Vecchia and  S. Ferrara, \np B130(1977)93;
 E. Witten, \pr D16(1977)2991.
\item {[31]} P.E. Haagensen, \ijmp A5(1990)1561;
 I. Jack and D.R.T. Jones, \pl B220(1989)176;
M.T. Grisaru and D. Zanon, \pl B184(1987)209.
\item {[32]} M.T. Grisaru, A. van de Ven and D. Zanon, \np B277(1986)409
M.T. Grisaru and D. Zanon, \pl B177(1986)347;
M.D. Freeman, C.N. Pope, M.F. Sohnius and K.S. Stelle, \pl
B178(1986)199;
Q-Han Park and D. Zanon, \pr D35(1987)4038.
\item {[33]}  S. Elitzur, A. Forge and E. Rabinovici, \np B359(1991)581;
 G. Mandal, A. Sengupta and S. Wadia, \mpl A6(1991)1685.
\item {[34]} C. Callan and Z. Gan, \np B277(1986)647;
 A.A. Tseytlin, \pl B241(1990)233; \pl B264(1991)311.
\item {[35]} J. Russo, L. Susskind and L. Thorlacius, preprint SU-ITP-92-24 .

\end